\documentclass[12pt]{article}
\usepackage{graphicx,amsbsy,latexsym,graphics}

\newwrite\ffile\global\newcount\figno \global\figno=1

\def\writedef#1{}

\input epsf
\def\figin{\epsfcheck\figin}\def\figins{\epsfcheck\figins}
\def\epsfcheck{\ifx\epsfbox\UnDeFiNeD
\message{(NO epsf.tex, FIGURES WILL BE IGNORED)}
\gdef\figin##1{\vskip2in}\gdef\figins##1{\hskip.5in}
\else\message{(FIGURES WILL BE INCLUDED)}%
\gdef\figin##1{##1}\gdef\figins##1{##1}\fi}

\def\figinsert{}
\def\ifig#1#2#3{\xdef#1{fig.~\the\figno}
\writedef{#1\leftbracket fig.\noexpand~\the\figno}%
\figinsert\figin{\centerline{#3}}\medskip\centerline{\vbox{\baselineskip12pt
\advance\hsize by -1truein\center\footnotesize{  Fig.~\the\figno.} #2}}
\bigskip\endinsert\global\advance\figno by1}
\def\endinsert{}

\begin{document}
\baselineskip 18pt
\newcommand{\Tr}{\mbox{Tr\,}}
\newcommand{\beq}{\begin{equation}}
\newcommand{\eeq}{\end{equation}}
\newcommand{\bea}{\begin{eqnarray}}
\newcommand{\eea}[1]{\label{#1}\end{eqnarray}}
\renewcommand{\Re}{\mbox{Re}\,}
\renewcommand{\Im}{\mbox{Im}\,}
\begin{titlepage}

\begin{picture}(0,0)(0,0)
\put(350,0){SHEP-02-08}
\end{picture}

\begin{center}
\hfill \vskip .4in {\large\bf ${\cal N}=4$ Super Yang Mills at Finite
Density : the Naked Truth}
\end{center}
\vskip .4in
\begin{center}
{\large Nick Evans and James Hockings
} \footnotetext{e-mail: evans@phys.soton.ac.uk,
jrh@hep.phys.soton.ac.uk} \vskip .1in
{\em Department of Physics, Southampton University,
Southampton, S017 1BJ, UK}
\end{center}
\vskip .4in
\begin{center} {\bf ABSTRACT} \end{center}
\begin{quotation}
\noindent We study ${\cal N}$=4 super Yang Mills theory at finite
U(1)$_R$ charge density (and temperature) using the AdS/CFT Correspondence.
The ten dimensional
backgrounds around spinning D3 brane configurations  split into
two classes of solution. One class describe spinning black branes
and have previously been extensively studied, and
interpreted, in a thermodynamic context, as
the deconfined high density phase of the dual field theory. 
The other class have naked singularities and in the
supersymmetric limit are known to correspond to multi-centre
solutions  describing the field theory in the Coulomb phase. We 
provide evidence
that the non-supersymmetric members of this class represent naked, spinning
D-brane distributions describing the 
Coulomb branch at finite density. At a critical density a phase transition
occurs to a spinning black brane representing 
the deconfined phase where the Higgs vevs have
evaporated. We perform a free energy calculation to
determine the phase diagram of the Coulomb branch 
at finite temperature and density.

\end{quotation}
\vfill
\end{titlepage}
\eject
\noindent

\section{Introduction}

In this letter we study ${\cal N}$=4 super Yang Mills theory at
finite U(1)$_R$ charge density using the AdS/CFT Correspondence
\cite{malda,gkp,w1}. The Correspondence allows us to 
study this phenomena non-perturbatively by investigating 
the background to a stack of spinning
D3 branes \cite{cvetic, larsen, jmc1}. The spin induces non-zero
components of the ten dimensional metric that after Kaluza Klein
reduction on the $S^5$ correspond to a vacuum expectation value
(vev) for a temporal U(1)$_R$ gauge boson. In the field theory
dual this field plays the role of a chemical potential putting the
field theory at finite density.

A set of metrics describing spinning D3 branes have been obtained by
oxidising five dimensional charged black hole solutions to
ten dimensions \cite{cvetic}. These ten dimensional solutions break
down into two classes. The first are rotating black branes and these
have been extensively studied in the literature \cite{cvetic, larsen, 
jmc1,spinbh}. In
terms of the duality with the ${\cal N}=4$ gauge theory they have been 
interpreted as the high density and high temperature deconfined phase.
The phase structure for the ${\cal N}=4$ theory at the origin of moduli
space was mapped out in \cite{jmc1,jmc2}. The second class of solutions 
are not black branes but nakedly singular metrics. The supersymmetric limit
of these solutions has been shown to correspond to disc distribution,
multi-centre D3
brane solutions \cite{larsen}. We will retell this story but using brane
probing techniques to motivate moving to coordinates in which the 
duality is manifest, in the spirit of \cite{ppb,jpe,beh,kahler}. 
Our main interest
though is in interpreting the non-supersymmetric members of this class.
Since they share many properties with the rotating black branes it
seems likely that they describe spinning disc distributions corresponding
to the Coulomb branch of the gauge theory at finite density. We provide
evidence that this is indeed the correct interpretation. 

The non-supersymmetric naked solutions only exist upto some maximum density
above which they develop a horizon and become the zero temperature
black brane solutions.
We interpret this transition, at which, as we will see,
the role of the parameters of the model radically change, as
the high density deconfinement transition of the Coulomb branch of the 
gauge theory above which the scalar vevs evaporate. 
We perform a free energy calculation comparing a
spinning D3 distribution background with a compact time dimension, and a black
brane geometry with the same temperature and density, to map out the phase 
diagram of the Coulomb branch. As one would expect the distribution size
(ie the size of the scalar vev) controls the transition temperature
and density. 

As has been noted elsewhere \cite{spinbh} for the black brane geometries, these
backgrounds are unstable if the spin or density is taken too large.
We show this using a brane probe computation (note probes of some 
related configurations were performed in \cite{op}). In fact the 
instability
of the gauge theory at zero temperature and finite density is readily
apparent perturbatively because the chemical potential destabilizes 
the scalar potential resulting in a run away vacuum \cite{superf}. The naive 
scalar instability in the theory is most clearly seen by observing 
that a probe D3 brane
in pure AdS$_5\times S^5$ experiences no potential and thus there
is no force to support rotational motion. Such a probe when given
angular momentum moves off to infinity corresponding to a runaway
scalar vev in the field theory. These geometries 
show that the instability remains non-perturbatively. Inspite of
this instability, the geometries nevertheless let us see the physics of the 
finite density phase transition.

\section{Finite Density}

The background we wish to study is the near horizon limit of a
rotating D3 brane configuration obtained by Cvetic et al
\cite{cvetic} from the lift of five dimensional charged black hole
solutions

\begin{eqnarray}
 ds^{2}_{10} & =  
& \sqrt{\tilde{\Delta}}\left[-\left(H_{1}H_{2}H_{3}\right)^{-2/3}f dt^{2}
            +
            \left(H_{1}H_{2}H_{3}\right)^{1/3}\left(f^{-1}dr^{2}+{r^{2}\over 
              L^2}
            d x_{//}^2\right)\right] \nonumber \\
         &    & + \frac{L^2}{ \sqrt{\tilde{\Delta}}}\sum^{3}_{i=1}X_{i}^{-1}\left(d\mu_{i}^{2}
                +\mu_{i}^{2}\left(d\phi_{i}+gA^{i}dt\right)^{2}\right) \label{label1}
\end{eqnarray}

\noindent where the $\mu_i$ are three direction cosines and

\beq f= - {\mu \over r^2} + {r^2 \over L^2} H_1 H_2 H_3, 
\hspace{1cm} {1 \over L^2}
= {1 \over \sqrt{2m} \sinh\alpha}, \hspace{1cm} \mu = {2 m \over L^2} \eeq

\beq  A^i = {L(1 - H_i^{-1}) \over l_i \sinh \alpha}dt, \hspace{1cm}
H_i = 1+{l_i^2\over r^2}\eeq

\beq \tilde{\Delta} = (H_1 H_2 H_3)^{1/3} \sum_i {\mu_i^2 \over
H_i}, \hspace{1cm} X_i = H_i^{-1} (H_1 H_2 H_3)^{1/3} \eeq

\beq B_{(4)} = - {r^4 \over L^4} H_1 H_2 H_3 \sum_i {\mu_i^2 \over H_i} dt
\wedge d^3x + {1 \over \sinh \alpha} ( \sum_i l_i \mu_i^2 d\phi_i) \wedge  
d^3x \eeq

At large $r$ the solution asymptotes to AdS$_5\times S^5$ with the
AdS radius $L$. We will keep $L$ fixed in the following analysis. 
The solution
then has four free parameters, the $l_i$ and $\mu$ (or equivalently $m$ or
$\alpha$ ).  For the  five dimensional black hole solutions the $l_i$
are charges under three U(1) gauge symmetries and $\mu$ the temperature.
We expect these parameters to lift to ten dimensions to be rotation
parameters in the three distinct U(1) planes of the $S^5$ and the temperature. 
The temperature of the black hole 
is given by

\beq 2 \pi T = \left.{1 \over \sqrt{g_{rr}}} {d \over d r}
\sqrt{g_{tt}}\right|_{r=r_h} \eeq 
where $r_h$ is the horizon
radius which can be determined from where the function $f=0$
\beq
r_H^4 H_1(r_H)H_2(r_H)H_3(r_H) = {\mu L^2} \label{label2}
\eeq
For
our ten dimensional solutions we find

\beq 4 \pi T = {2 \over L^2} r_H (H_1H_2H_3)^{1/2} 
\left( 2 - {1 \over L \mu^{1/2} 
(H_1H_2H_3)^{1/2}} (l_1^2 H_2H_3 + l_2^2 H_1 H_3 + l_3^2 H_1 H_2) \right) \label{label4}
\eeq 

We can solve these equations for a number of special cases to find the
value of $\mu$ that corresponds to $T=0$. For example when a single $l_i$
is non-zero $T=0$ corresponds to $\mu=0$, for two equal non-zero $l_i$
$T=0$ corresponds to $\mu = l^4/L^2$ and when all three $l_i$
are equal $T=0$ corresponds to $\mu = 27 l^4/4L^2$. For $\mu$ equal
to these $\mu_c$ values and above the solutions have a singularity, 
orginating in the $f$ function, which corresponds to the horizon of 
the black hole. As $\mu$ increases the black hole temperature increases.
However, for $0 < \mu < \mu_c$ the solutions do not have a horizon
but have a naked singularity at $r=0$.
We show this in the plots of figure 1 where $g_{rr}$ is plotted 
against $r$ at varying $\mu$ for the case when all three $l_i$ are equal.

The black hole/brane solutions are closely related to those analysed in
\cite{jmc1,jmc2} to describe the behaviour of ${\cal N}$=4 super Yang Mills
at finite temperature and density. In \cite{jmc1,jmc2} the three $l_i$ were
taken equal and the variant
of the above metric where the Minkowski space slices of AdS are
compactified on $S^3$ was considered\footnote{Placing the 
gauge theory on $S^3$ introduces an extra scale into the problem
which enlarges the region of the thermodynamic temperature
vs density plane where the confined phase survives \cite{jmc1}. 
On $R^3$
the phase transition to the deconfined phase occurs the 
moment that a temperature or chemical potenital is introduced}. 
The parameters $l_i$ control the
rotation speed of the black hole or the chemical potential in the
field theory. 
The parameter $\mu$ controls the temperature of the
black hole (or in the dual field theory) 
with $\mu = \mu_c$ corresponding to T=0.
Following \cite{jmc1,jmc2} these black hole solutions 
should be interpreted as
gravity duals of the field theory at the origin of moduli space 
across the full temperature and
density plane (the origin is described by the usual AdS/CFT
correspondence). The behaviour of a Wilson loop \cite{w2} in these
backgrounds show that at finite chemical potential and temperature
the theory lives in a distinct (deconfined) phase from the
(confined) theory at the origin.

\begin{figure}
\begin{center}
\hskip-10pt{\lower15pt\hbox{
\epsfysize=2.5 truein \epsfbox{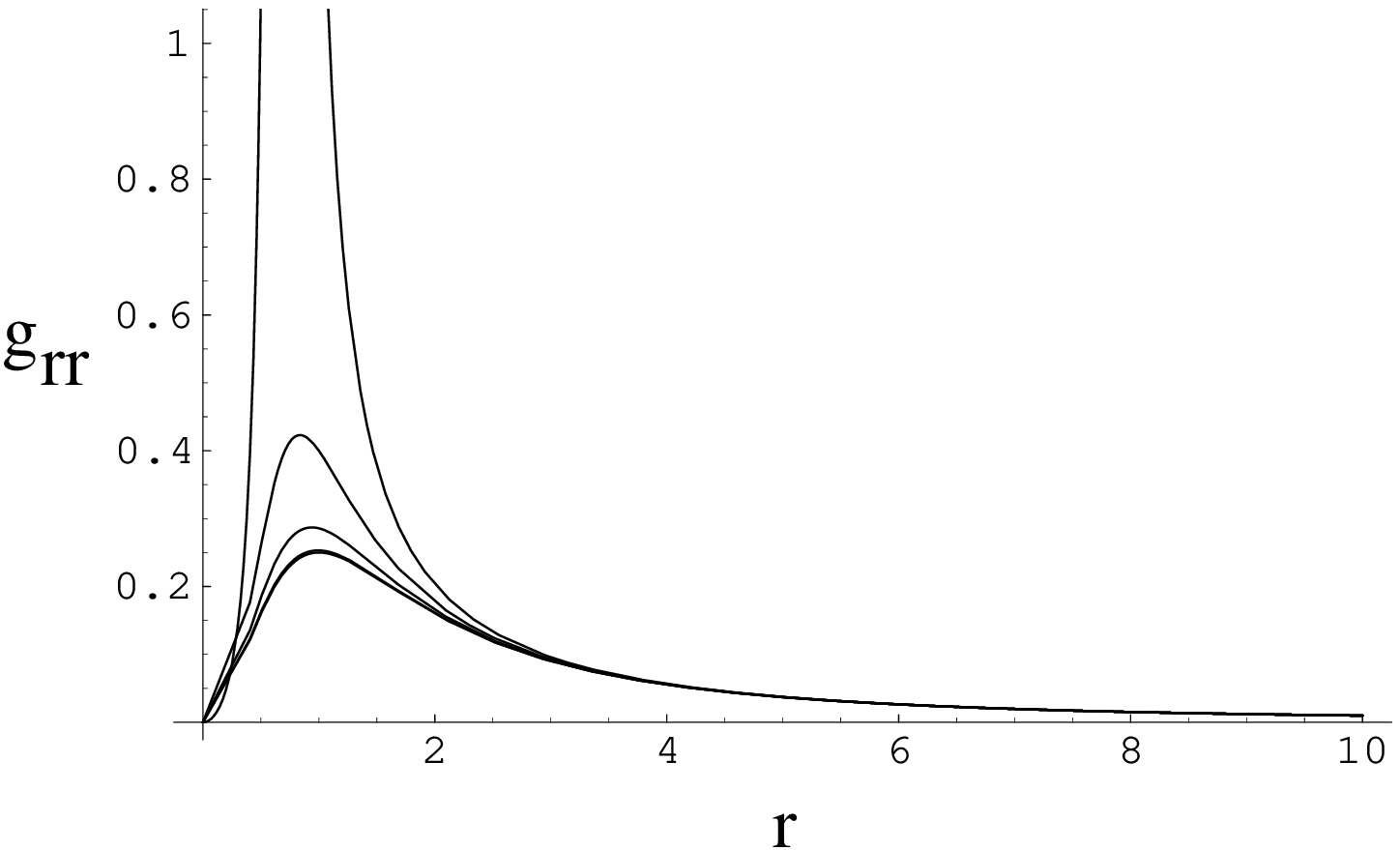}}} \medskip

Figure 1: The metric component $g_{rr}$ plotted as a function of
radial position for varying $\mu$ up to $\mu = \mu_c$ at
fixed $l_1=l_2=l_3=l$ showing the development of a horizon. 
\end{center}
\end{figure}

\subsection{Spinning Discs}

What role then is there for the nakedly singular solutions when
$\mu < \mu_c$? Traditional thinking would declare these
backgrounds unphysical, however, recent developments have shown
that naked singularities need not be pathological \cite{gub2,pjp}
but may simply represent the presence of extended objects, such as
D-branes, in the space. The most straight forward examples of such
a background are the multi-centre solutions \cite{larsen,
freed2, brr} describing a distribution of D3 branes which provide a
dual description of the Coulomb branch of the ${\cal N}$=4 gauge
theory. In fact it has already been shown in \cite{larsen} that
the, supersymmetric, $\mu \rightarrow 0$ limit of precisely the
singular backgrounds we consider here are multi-centre solutions.
We begin with that analysis and will then consider what happens as
$\mu$ is switched on.

One must be careful in taking the $\mu \rightarrow 0$ limit to
remember to keep $L$ fixed which also requires $\alpha \rightarrow
\infty$. The background becomes
\begin{eqnarray}
 ds^{2}_{10} & =  & \sqrt{\tilde{\Delta}}\left[\left(H_{1}H_{2}H_{3}\right)^{1/3}{ r^2 \over L^2}
              (-dt^{2}+ dx_{//}^2)            +
            {L^2 (H_{1}H_{2}H_{3})^{-2/3} \over r^2} dr^{2}
            \right]\\
\nonumber   &&    +  
\frac{L^2}{ \sqrt{\tilde{\Delta}}}\sum^{3}_{i=1}X_{i}^{-1}\left(d\mu_{i}^{2}
                +\mu_{i}^{2}\left(d\phi_{i}\right)^{2}\right)
\end{eqnarray}

\beq B_4 = - {r^4 \over L^4} H_1 H_2 H_3 \sum_i {\mu_i^2 \over H_i} 
dt \wedge dx^3 \eeq

Note that the one form potential vanishes in this limit leaving a
non-rotating solution.  The difficulty with interpreting
backgrounds as duals of gauge theory is though the familiar
problem of finding the coordinates appropriate to the duality.
Brane probing has proven itself to be an especially useful tool in
this respect since it converts the background to the abelian gauge
theory on the world volume of the probe where we can use field
theory intuition to find the correct coordinates \cite{ppb,jpe,
beh,kahler}. Thus we place a slow moving D3 brane in the background
through the Born Infeld action

\beq S = -{\tau_3 \over g_s} \int d^4 \xi \sqrt{- det g_{ab}} - \mu_3 \int B_4
\eeq where $\tau_3 = \mu_3 g_s^{-1}$ and $g_{ab}$ is the pull back of the
background to the world sheet. We find the action

\beq L = {1 \over 2} \left( \sum_i {\mu_i^2 \over H_i^2} \dot{r}^2
+ \sum_i {1 \over 2} r^2 H_i (\dot{\mu}_i^2 + \mu_i^2
\dot{\phi}_i^2) \right) \label{pot} \eeq

There is no potential obstructing motion of the probe in the six
dimensional transverse space giving a strong hint that the
theory is indeed the pure ${\cal N}$=4 theory. In the coordinates
appropriate to the duality we expect a canonical kinetic term for
the six scalar fields on the probe suggesting we try the new
coordinates 

\beq \label{coord} w^2 \tilde{\mu}_i^2 = (r^2 + l_i^2) \mu_i^2 \eeq
which render the $\dot{\phi}^2$ terms canonical. It follows that

\beq w^2 = \sum_i (r^2 + l_i^2) \mu_i^2 \eeq

These are the coordinates identified in \cite{larsen} that convert
the metric to the familiar form of a multi-centre solution. 
They transform the probe action
so that it has a flat metric on moduli space
and leave the spacetime background in the form 

\beq ds_{10}^2 = H_D^{-1/2} dx_{//}^2 + H_D^{1/2} dw^2 \eeq
with
\beq B_4 = - H_D^{-1}  dt \wedge dx^3 \eeq

We may find the form of $H_D$ from the $g_{xx}$ component of the metric
using the coordinate transformation in (\ref{coord}). For example, for
a single $l_i$ switched on we find

\beq
H_D^{-1} =  {1 \over L^4}( w^2 - l^2 \mu_1^2 )^2 \left( \mu_1^2 + H \mu_2^2 +
H \mu_3^2 \right), \hspace{1cm} H = 1 + {l^2 \over w^2 -l^2 \mu_1^2} \eeq
where

\beq \mu_1^2 = {w^2 \over w^2 + l^2(1-\mu_1^2)} \tilde{\mu}_1^2,
\hspace{1cm} \mu_{2/3}^2 = {w^2 \over w^2 - l^2\mu_1^2}
\tilde{\mu}_{2/3}^2  \eeq

and thus

\beq \mu_1^2 = {(w^2+l^2) \pm \sqrt{(w^2+l^2)^2-4l^2w^2
\tilde{\mu}_1^2 }\over 2 l^2} \eeq

The result is unenlightening, except that if we look 
in the $\phi_1$  plane at  $w=l$ 
by setting $\tilde{\mu}_1=1, \tilde{\mu}_{2/3}=0$ which
corresponds to $\mu_1=1, \mu_{2/3}=0$ at $w=l$ and we find $H_D=0$.
The metric in this case is singular at $w=l$, or in the original
coordinates $r=0$. The singularity corresponds to the position of the 
D3 brane distribution responsible for the background - it is a disc in the 
$\phi_1$ plane at $w=l$. 

Similar manipulations for the case with two equal $l_i$ give

\beq H_D^{-1} =  {1 \over L^4}( w^2 - l^2 (\mu_1^2 + \mu_2^2) )^2 
\left( \mu_1^2 + \mu_2^2 +
H \mu_3^2 \right), \hspace{1cm} H = 1 + {l^2 \over w^2 -l^2 (\mu_1^2
+\mu_2^2)} \eeq
Again looking at $w=l$ and setting $\tilde{\mu}_3=0$ ($\mu_3=0$) so
$\tilde{\mu_1}+\tilde{\mu_2}=1$ ($\mu_1+\mu_2=1$) we find 
singularities in the four dimensional space described
by the $\phi_1$ and $\phi_2$ planes corresponding to a spherical D3
distribution in that space. The case with three equal $l_i$ gives
the much simpler result

\beq H_D^{-1} = w^4 /L^4 \eeq
Here the distribution is an S$^5$ at $w=l$ as can be deduced from 
the fact that the $r$ coordinates only extend to $r=0$ or $w=l$ or
by following the deformation of one of the above singular distribution 
as $l_3$ is switched on. Note that the $S^5$ distribution does not show up
as singularities in $H_D$ because it is an SO(6) singlet and
hence does not appear in the supergravity because it is not an
operator in a short multiplet. The space is AdS$_5 \times S^5$ truncated
at $w=l$.

Now we have identified the physical coordinates for $\mu=0$ we can
consider turning $\mu$ back on for a fixed distribution (fixed $l_i$). 
Turning on  $\mu$ introduces spin or  
finite density in the 
field theory as can be seen by looking at the metrics at large $w$ 
($\simeq r$) where they
look like AdS with a gauge potential

\beq A_i \simeq {l_i \mu^{1/2} \over r^2} \eeq 
In this limit we may treat the solution as a five dimensional solution
and  calculate the U(1)$_R$ charge in the interior.
We can thus deduce a charge density in the dual field theory 
associated with each of the three U(1)$_R$ 
subgroups of SU(4)$_R$ which are 
proportional to $l_i \mu^{1/2}$. It seems reasonable to conclude that 
we are observing a solution describing a spinning version of the 
disc distribution.

We should be careful to check for evidence that no other deformations
of the theory have occured.
Let us first address this issue in the 
middle example above where two of the $l_i$ are switched on with equal 
values. As $\mu$ switches on note that the $g_{xx}$ component of the metric
is unchanged by the inclusion of $\mu$. The singularity locus in this
component remains at the same place. Also the four
form $dt \wedge dw^3$ piece is unchanged showing that the number
of D3 branes in the interior is unchanged. 
Finally we note that $\mu$ introduces
no angular dependence in the $\phi_1$ or $\phi_2$ plane. The conclusion of
these facts is that $\mu$ does not change the angularly
constant in the $\phi_1$ and $\phi_2$ planes, 
distribution of D3 branes at $w=l$.

Thus the metrics with $\mu < \mu_c$ seem to naturally
describe spinning versions of the multi-centre solution
corresponding to the dual ${\cal N}$=4 theory being on its coulomb
branch with a chemical potential. In fact it is clear that these 
metrics must describe such configurations because they are the unique
solutions of the field equations with the symmetries of these systems
(it is particularly clear that a spinning $S^5$ distribution of
D3 branes will share the symmetries of a spinning black hole).
This sharing of symmetries between the black hole solutions and rotating
D-brane distributions explains why the two sets of solutions are
naturally intertwined.

\subsection{Finite Density Phase Transition}

It is interesting that we
can not increase the chemical potential to infinity for a fixed
distribution (fixed $l_i$) and maintain 
a rotating distribution form for the solution - at $\mu =\mu_c$ 
there is a transition to a black brane and we lose all
information about the interior structure. In the field theory at
this critical density apparently knowledge of the scalar vevs is
lost. Note that there is a sharp change in the interpretation of
the parameters of the solution. When the interior is naked the
solution must provide information about the interior structure
which it does through the parameters $l_i$ and then $\mu$ plays the
role of rotation speed. Above the critical $\mu$ there is a black
brane and knowledge of the interior structure is lost and so $l_i$
switch to describing the rotation and $\mu$ describes the newly
available parameter, temperature.

In the field theory dual we must be seeing the {\it finite density
transition of the coulomb branch} where the scalar potential is
forced to favour zero vevs. When the chemical potential
is much less than the scalar vevs the vevs will be unaffected
whilst when the chemical potential is much larger the theory should 
look like the deconfined phase of the theory at the origin of 
moduli space. The scale of the transition should be set by
the size of the vevs ($l_i$) and indeed we have seen $\mu_c \sim l^4$.
Above the critical density the
spacetime is a black hole, a phase that has been identified with
the deconfined phase of the field theory, as we would expect for
the phase when the scalar vevs evaporate. 

If we begin with a black brane metric with $\mu=\mu_c$ ($T=0$)
and want to decrease the chemical potential we now realize there
are two possibilities in the field theory. If the theory has small
or zero vev it will remain in the deconfined phase as we decrease
the density, else, if the theory has a large vev, then as we
decrease the density below that vev the system should undergo a
transition to the Coulomb phase. It's now clear that the dual
background elegantly offers us both of these choices! We can
decrease the density in two ways - either we keep $\mu=\mu_c$ and
decrease $l$ in which case we retain a black brane configuration
corresponding to the first case in the field theory, or we can
keep $l$ fixed and decrease $\mu$ in which case we obtain a
spinning multi-centre solution describing the coulomb phase.

The solutions with the three $l_i$ equal fit this story equally well except 
that there is no singularity to monitor the position of the D3 branes
as $\mu$ is switched on. Again by considering deformations of other
singular configurations it is clear that the interpretation is 
the same as that just given. The metrics with a single $l_i$ switched
on, however, do not show this behaviour. In fact as we saw above the 
condition for a T=0 black brane is precisely $\mu=0$ where the solution 
becomes a supersymmetric non-rotating disc distribution. For some reason
these metrics do not provide us with {\it any} description of the rotating 
zero temperature states. Presumably this is just a failure of the completeness
of these solutions rather than anything more subtle and we would expect 
similar behaviour on that part of the coulomb branch if only we had the
appropriate metrics. 

We note that this transition from the Coulomb phase to the
deconfined phase is also apparent in the similar solutions in
which the Minkowski space slices of AdS are compactified
\cite{jmc1,jmc2}. Recently Myers and Tafjord \cite{myta} have
argued that the nakedly singular metrics in that case correspond
to distributions of giant gravitons. Again though above some
critical angular momentum the solutions shift to black hole
solutions showing that at high enough density the giant gravitons
are forced to evaporate leaving a deconfined phase.

\subsection{Stability}

Many authors \cite{spinbh,op} 
have studied the stability of the black brane solutions
within the class of geometries under discussion 
and concluded that they are unstable for large densities.
This is to be expected \cite{superf} 
since if, at zero temperature, we introduce a
chemical potential into the ${\cal N}$=4 gauge theory at tree
level via a vev for the temporal component of a spurious U(1)$_R$
gauge field then there will be a contribution to the scalar
potential since the scalars are in the 6 of SU(4)$_R$.

\beq \Delta L = | D^\mu \phi | \rightarrow A_0^2 |\phi|^2 \eeq
A negative mass term is introduced for the scalar which will
destabilize the moduli space of the theory, giving rise to a
runaway potential. The same phenomena is apparent if we try to
introduce rotation for a D3 probe in AdS space. Since there is no
potential in the transverse space (as we saw in (\ref{pot})
above), rotational motion can not be supported and the D3 brane
will progress to the edge of the moduli space displaying the
runaway scalar vev. This argument is of course naive because
quantum effects could stabilize the potential. The backgrounds to
spinning D3 branes discussed above though provide a complete dual
description of the field theory with a chemical potential and we
may determine their stability by finding the potential seen by a
probe in their background. As an example the resulting probe potential in the
case where the three $l_i$ are set equal, and its
expansion for small $A_0 = A_i(w=l) = \mu^{1/2}/l$ is

\beq  V = {w^4 \over L^4} \left( 1 - \sqrt{1 - {L^2 A_0^2 l^2 \over w^4}}
\right)\simeq {1 \over 2L^2}  A_0^2 l^2 + {1 \over 8}{A_0^4 l^4
\over w^4} + ...\eeq

The probe is forced to infinity by the potential (we plot the full
expression in figure 2). We deduce that the whole configuration is
indeed unstable since any of the D3s in the distribution can be
considered as the probe.
Remarkably, these backgrounds have though allowed us to explore
the finite density behaviour of the Coulomb branch of the theory
ignoring this instability.

\begin{figure}
\begin{center}
\hskip-10pt{\lower15pt\hbox{
\epsfysize=2.5 truein \epsfbox{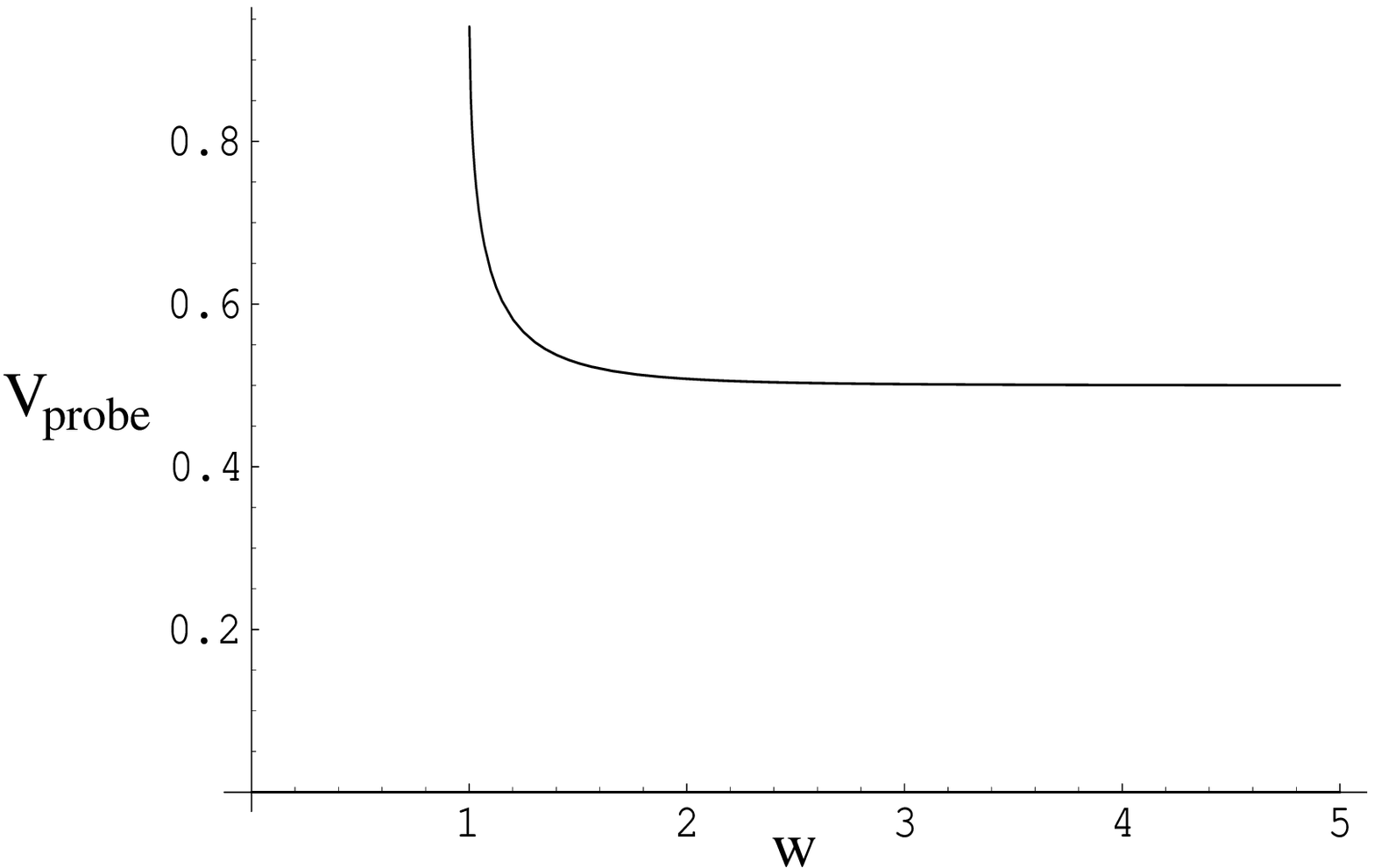}}} \medskip

Figure 2: The form of the probe potential as a function of radial distance in
the spinning D3 background.
\end{center}
\end{figure}

\section{Thermodynamics of the Coulomb Branch}

We will now extend our analysis to include finite temperature and map out 
the phase diagram of a point on the Coulomb branch,
in the spirit of the Hawking Page transition \cite{hawkpage}. 
For ease of calculation
we will study points on the Coulomb branch where the global $SU(4)_R$
symmetry is preserved; these are distributions in which the D3 branes live 
on an $S^5$. We will therefore study the naked solutions above with all
three $l_i$ equal and fixed. These solutions exist upto $\mu_c$ and have been
identified above with an $S^5$ distribution of D3 branes spinning equally
in the three transverse planes. To study these solutions at non-zero
temperature we must compactify the time like direction with period $\beta 
= 1/T$. The chemical potential of these geometries is given by $\mu^{1/2}
l$. The full set of geometries above also contain black brane solutions
with the same temperature and chemical potential (above the chemical potential 
value $\mu_c^{1/2} l$ there are only black brane solutions so we study the
parameter space below that point). To find which of these
solutions is the energetically preferred solution we must calculate the
free energy difference. The appropriate action is

\beq I = -{1 \over 2\kappa^2} \int d^{10}x \sqrt{-G}\left(R + {1 \over 480}G^2_{(5)}\right)
	 -{1 \over \kappa^2} \int d^{9}x \sqrt{-h_{(9)}}\overline{K} \eeq

The second integral is a surface term where

\beq \overline{K} = G^{\mu\nu}K_{\mu\nu}, \hspace{1cm}
 K_{\mu\nu} = {1 \over 2\sqrt{G_{rr}}}{\partial \over \partial r}G_{\mu\nu}, 
\hspace{1cm} h_{(9)} = det{G_{\mu\nu}}, \hspace{1cm} \mu,\nu \neq r \eeq

As described in \cite{hawkpage,w2}, to allow comparison of the two spacetimes 
the period of the time integral of the naked solution, 
$\tilde{\beta}$, must be 
set to match the geometry of the hypersurface at large $R$ in the two cases. 
To achieve this we require

\beq \tilde{\beta} = \beta {\sqrt{G_{tt}} \over \sqrt{\tilde{G_{tt}}}} \eeq

For the metrics under consideration calculation shows that 
the curvature, $R=0$, leaving us with just the five-form and surface pieces.
We use subscripts on the $\mu$ and $l$ parameters to distinguish the naked and 
black brane cases; a $1$ subscript denotes the black brane and a $2$ 
the naked geometry. Direct computation gives the action difference 

\beq
I = I_1 - I_2 = {1 \over \kappa^2} Vol(S_5)Vol(3)\beta 
\left( 2l_2^4 - 2l_1^4 -
	\mu_2 + \mu_1 - 2r_h^4 - 4r_h^2l_1^2 \right) \label{label3}
\eeq
where $Vol(S_5)$ is the volume element associated with the angular integration
which is common to the two geometries, $Vol(3)$ is the 
volume of the spatial part of the branes and 
$r_h$ is the horizon radius of the 
black hole. 

\begin{figure}
\begin{center}
\hskip-10pt{\lower15pt\hbox{
\epsfysize=2.5 truein \epsfbox{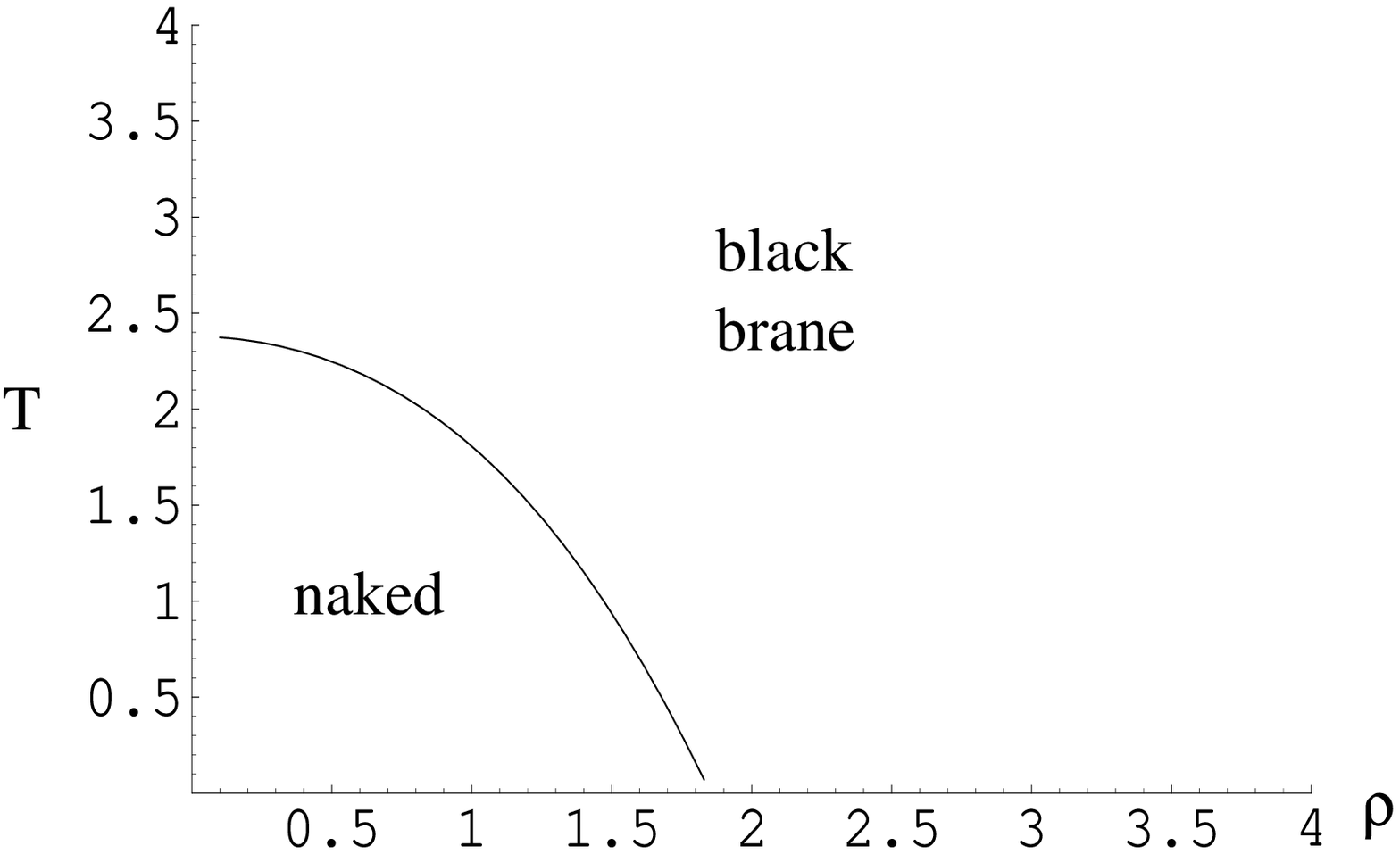}}} \medskip

Figure 3: The temperature-density plane, showing the critical line inside 
which one has the naked solutions (Coulomb phase), 
and beyond it the black branes 
(deconfined phase) \end{center}
\end{figure}

The action calculation indeed reveals a phase transition between the 
two geomtries as a function of temperature and chemical potential.
For low temperature and density the naked solution is preferred whilst
for high temperature and density the black brane solution
is preferred. The transition occurs essentially when the black brane radius
becomes larger than the distribution size as expected, since these are
the only two scales in the problem. We note that the transition 
on the zero temperature axis actually occurs  a little below the value $\mu_c = 27/4 l^4$ determined earlier. 
Thermodynamically there is no reason why this shouldn't be true - the phase
diagram still matches expectations. 
It does though make the precise interpretation
of $\mu_c$ more opaque. 
In the dual field theory at low temeprature
and density the solution describes a point on the Coulomb branch with
scalar vevs. At high temperature and density the theory transitions
to a deconfined phase without scalar vevs. The transition occurs
when the temperature or chemical potential is of order the scalar vevs.
In figure 3 we plot the form of the phase diagram where it can be seen 
that the result of the supergravity calculation matches field theory
expectations. \vspace{1in}

\noindent {\bf Acknowledgements}\vskip .1in \noindent NE is grateful for
the support of a PPARC Advanced Fellowship. JH is grateful for the
support of the Department of Education, Isle of Man and his
parents. The authors are grateful to Clifford Johnson for first alerting
them to the existence of the naked solutions and allowing them access to 
his notes on black hole thermodynamics.


\begin{thebibliography}{6666666666}
%
\bibitem{malda} J. Maldacena, Adv. Theor. Math. Phys. 2 (1998) 231,
hep-th/9711200.
%
\bibitem{gkp} S.S. Gubser, I.R. Klebanov and A.M. Polyakov,
Phys. Lett. B428 (1998) 105, hep-th/9802109.
%
\bibitem{w1} E. Witten, Adv. Theor. Math. Phys. 2 (1998) 253, hep-th/9802150.
%
\bibitem{cvetic} M. Cvetic, M.J. Duff, P. Hoxha, J.T. Liu, H. Lu, J.X. Lu, 
R. Martinez-Acosta, C.N. Pope, H. Sati, T.A. Tran, Nucl. Phys. {\bf B558}
(1999) 96, hep-th/9903214.
%
\bibitem{larsen} P. Krauss, F. Larsen, S.P. Trivedi, JHEP {\bf
9903} (1999) 03; hep-th/9811120.
%
\bibitem{jmc1} A. Chamblin, R. Emparan, C.V. Johnson, R.C. Myers,
Phys. Rev. {\bf D60} (1999) 064018, hep-th/9902170. 
%
\bibitem{jmc2} A. Chamblin, R. Emparan, C.V. Johnson, R.C. Myers,
Phys.Rev. {\bf D60} (1999) 104026, hep-th/9904197.
%
\bibitem{spinbh} S.S. Gubser,  Nucl. Phys. {\bf B551} (1999) 667, 
hep-th/9810225; R. Cai, K. Soh, Mod. Phys. Lett. {\bf A14} (1999) 1895,
hep-th/9812121; M. Cvetic, S.S. Gubser, JHEP {\bf 9907} (1999) 010, 
hep-th/9903132; T. Harmark, N.A. Obers, JHEP {\bf 0001} (2000) 008,
hep-th/9910036; J.G. Russo, K. Sfetsos, Adv. Theor. Math. Phys. {\bf 3}
(1999) 131, hep-th/9901056
%
\bibitem{ppb} A. Buchel, A. Peet, J. Polchinksi, Phys. Rev. {\bf D63} 
(2001) 044009, hep-th/0008076.
%
\bibitem{jpe} N. Evans, C.V. Johnson, M. Petrini, JHEP {\bf 0010} (2000)
022, hep-th/0008081.
%
\bibitem{beh} J. Babington, N. Evans, J. Hockings, JHEP {\bf 0107} (2001) 034,
 hep-th/0105235.
%
\bibitem{kahler} C.V. Johnson, K.J. Lovis, D.C. Page, JHEP {\bf 0110} 
(2001) 014, hep-th/0107261.
%
\bibitem{op} R. Cai,  JHEP {\bf 9909} (1999) 027, hep-th/9909077.
%
\bibitem{superf} N. Evans and M. Petrini, JHEP {\bf 0111} (2001) 043, 
hep-th/0108052.
%
\bibitem{gub2} S.S. Gubser, Adv. Theor. Math. Phys. {\bf 4} (2002) 679, 
hep-th/0002160.
%
\bibitem{pjp} C. V. Johnson, A. W. Peet and J. Polchinski,
Phys. Rev. D61 (2000) 086001, hep-th/9911161.
%
\bibitem{freed2} D. Z. Freedman, S. S. Gubser, K. Pilch and
N. P. Warner, JHEP {\bf 0007} (2000) 038, hep-th/9906194.
%
\bibitem{brr}A. Brandhuber and K. Sfetsos, Adv. Theor. Math. Phys.
{\bf 3} (1999) 851, hep-th/9906201; I. Chepelev and R. Roiban,
Phys.Lett. B462 (1999) 74, hep-th/9906224. I. Bakas and K.
Sfetsos, Nucl.Phys. B573 (2000) 768, hep-th/9909041.
%
\bibitem{w2} E. Witten, Adv. Theor. Math. Phys. 2 (1998) 505, hep-th/9803131;
J.P. Gregory, S.F. Ross, Phys. Rev. {\bf D63} (2001) 104023, hep-th/0012135. 
%
\bibitem{myta} R.C. Myers, O. Tafjord,
JHEP {\bf 0111} (2001) 009,  hep-th/0109127.
%
\bibitem{hawkpage} S.W. Hawking, D.N. Page, Commun. Math. Phys. {\bf 87}
(1983) 577.
%

\end{thebibliography}
\end{document}